\documentclass{synmet}

\usepackage{epsfig}

 \usepackage{amssymb}

 \advance\topmargin by -18mm

\begin{document}

\begin{frontmatter}


  
  \title{Comparison of experimental data and theoretical
    calculations for electrical resistivity and Hall coefficient in
    $\rm(TMTSF)_2PF_6$}


\author[label1]{Victor M.~Yakovenko\thanksref{label3}}
\author[label2]{and Anatoley T.~Zheleznyak}
\thanks[label3]{Corresponding author. 
Phone:  +1-301-405-6151,
Fax:    +1-301-314-9465,
E-mail: yakovenk@physics.umd.edu,
Web:    http://www2.physics.umd.edu/\~{}yakovenk/
}

\address[label1]{Department of Physics and Center for
  Superconductivity Research, University of Maryland, College Park, MD
  20742, USA}

\address[label2]{MetricVision, a Thermo Electron Company, 8500 Cinder
  Bed Road, Suite 150, Newington, VA 22122, USA}

\vspace{-\baselineskip}

\begin{abstract}
  The temperature dependences of the Hall coefficient and electrical
  resistivity recently measured by Moser {\it et al.} [Phys. Rev.
  Lett.  {\bf 84}, 2674 (2000)] in the quasi-one-dimensional organic
  conductor $\rm(TMTSF)_2PF_6$ are quantitatively compared with our
  previous theoretical calculations [Synth. Met. {\bf 103}, 2202
  (1999); Eur. Phys. J. B {\bf 11}, 385 (1999)].  We find a good
  agreement, albeit not with a fully consistent set of parameters for
  the two quantities.  \vspace{-\baselineskip}
\end{abstract}

\end{frontmatter}

Recently, the temperature dependences of the Hall coefficient, as well
as electrical resistivity, were measured in the quasi-one-dimensional
(Q1D) organic conductor $\rm(TMTSF)_2PF_6$ by the two groups
\cite{Jerome,Forro}.  The results were interpreted using the Luttinger
liquid concept in Ref.\ \cite{Jerome} and the conventional Fermi
liquid theory in Ref.\ \cite{Forro}.  Before the experimental data
became known, we had published theoretical calculations of the
temperature dependences of the Hall coefficient \cite{R_H} and
electrical resistivity \cite{R_xx}.  In the present paper, we
quantitatively compare our theoretical predictions and the
experimental results.  Since we have calculated the Hall coefficient
in the (a-b) plane, we compare only with the results of Ref.\ 
\cite{Jerome}, where the Hall coefficient was measured in that
geometry, unlike in Ref.\ \cite{Forro}, were the (b-c) plane geometry
was employed.  Electrical resistivity, calculated in Ref.\ \cite{R_xx}
for the direction along the chains, is compared only with the
experimental results of Ref.\ \cite{Jerome}, because they correspond
to constant volume, unlike the constant-pressure results of Ref.\ 
\cite{Forro}.

The Hall coefficient $R_H=\sigma_{xy}/H\sigma_{xx}\sigma_{yy}$, where
$\sigma_{ij}$ are the components of the conductivity tensor and $H$ is
the magnetic field, is usually a constant, because, in a simple Drude
model, $\sigma_{xy}\propto\tau^2$ and $\sigma_{xx,yy}\propto\tau$, so
the temperature-dependent relaxation time $\tau$ cancels out in $R_H$.
However, the experiment \cite{Jerome} found that $R_H$ in
$\rm(TMTSF)_2PF_6$ does depend on temperature.  The authors invoked
the Luttinger liquid concept in order to explain this effect.
However, it has been shown theoretically that $R_H$ of weakly coupled
one-dimensional Luttinger chains does not depend on frequency and
temperature \cite{Lopatin}, because the power-law dependences of
$\sigma_{xy}$ and $\sigma_{yy}$ cancel out.

\begin{figure}[t]
\centerline{\psfig{file=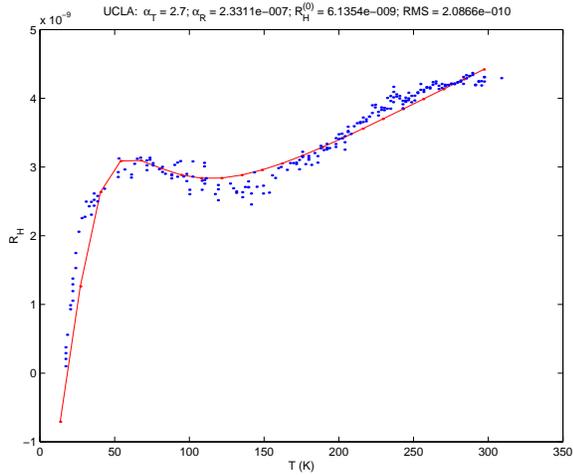,width=\linewidth}}
\caption{Temperature dependence of the Hall coefficient $R_H$
  in the units of $m^3/C$.  Experimental points are from Fig.\ 1 of
  Ref.\ \cite{Jerome} for the UCLA sample.  The solid line is the
  theoretical curve (a) from Fig.\ 1 of Ref.\ \cite{R_H}.}
\label{fig:R_H}
\end{figure}

The Hall coefficient may depend on temperature if the system has
relaxation times with different temperature dependences.  (See
discussion \cite{YBCO} of the ``cold spots'' for the cuprate
high-temperature superconductors, where $R_H$ is also
temperature-dependent.)  In Ref.\ \cite{hotspots}, we had calculated
the distribution of the umklapp scattering rates over the Fermi
surface of a Q1D conductor and found that at low temperatures ``hot
spots'' develop with a different temperature dependence.  Using this
distribution, we have calculated the temperature dependence of $R_H$
in Ref.\ \cite{R_H}.  Fig.\ \ref{fig:R_H} shows our theoretical curve
with the experimental points from Ref.\ \cite{Jerome}.  In our
calculation, the Hall coefficient consists of two terms:
$R_H=R_H^{(0)}+R_H^{(1)}$.  The first term is a
temperature-independent band-structure contribution.  In our fit, it
is taken to be $R_H^{(0)}=6.1\times10^{-9}\:\rm m^3/C$, which is 1.7
times greater than the value assumed in Ref.\ \cite{Jerome}.  The
second, temperature-dependent term $R_H^{(1)}$ is obtained by
multiplying curve (a) of Fig.\ 1 from Ref.\ \cite{R_H} by the factor
$\alpha_R=2.3\times10^{-7}\:\rm m^3/C$.  That corresponds to
$p_Fv_F/t_b=38$, which is 0.57 of the value assumed in Ref.\ 
\cite{R_H}.  The temperature scale of Fig.\ 1 from Ref.\ \cite{R_H} is
increased by the factor $\alpha_T=2.7$, which means that the
transverse tunneling amplitudes $t_b$ and $t_b'$ are taken to be
$\alpha_T$ times greater than the values 300 K and 30 K assumed in
Ref.\ \cite{R_H}.  Line (a) in Fig.\ 1 of Ref.\ \cite{R_H} corresponds
to the phases $\varphi=\varphi'=0$ in the transverse electron
dispersion law (see the definitions in Ref.\ \cite{R_H}).  The other
curves shown in Fig.\ 1 of Ref.\ \cite{R_H} for different values of
the phases qualitatively disagree with the experimental behavior
\cite{Jerome}.

\begin{figure}[t]
\centerline{\psfig{file=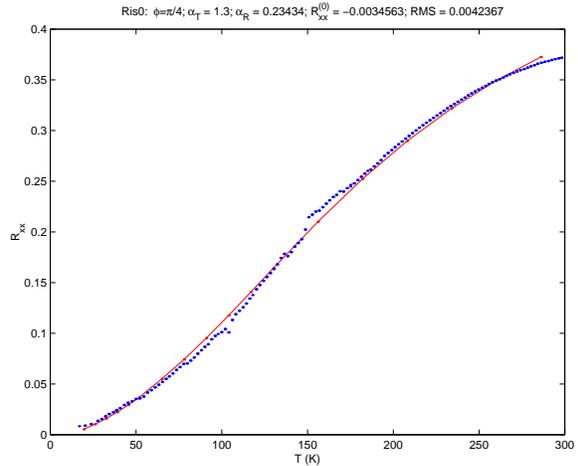,width=\linewidth}}
\caption{Temperature dependence of electrical resistivity along  the
  chains.  Experimental points are from Fig.\ 2 of Ref.\ \cite{Jerome}
  for the Ris{\o} sample.  The solid line is the theoretical curve
  with $\varphi=\pi/4$ and $t_b'=20$ K from Fig.\ 4 of Ref.\ 
  \cite{R_xx}.}
\label{fig:R_xx}
\end{figure}

In Ref.\ \cite{R_xx} we have calculated the temperature dependence of
electrical resistivity $R_{xx}$ along the chains by taking into
account renormalization of umklapp scattering.  Fig.\ \ref{fig:R_xx}
shows our theoretical curve with the experimental points from Ref.\ 
\cite{Jerome}.  The theoretical line is taken from Fig. 4 of Ref.\ 
\cite{R_xx} with the parameters $\varphi=\pi/4$, $\varphi'=2\varphi$,
$t_b=290$ K, $t_b'=$ 20 K, and the temperature scale is increased by
the factor $\alpha_T=1.3$.  The vertical scale of Fig. 4 of Ref.\ 
\cite{R_xx} is multiplied by the factor $\alpha_R=0.23$.

Overall, the theory correctly captures the experimental features.
However, the sets of model parameters utilized to obtain the fits in
Figs.\ \ref{fig:R_H} and \ref{fig:R_xx} are not the same.  First, the
values of the phases $\varphi$ and $\varphi'$ are different.  The Hall
coefficient is very sensitive to the choice of $\varphi$ and
$\varphi'$, and a sensible fit is possible only when the phases are
zero or close to zero.  On the other hand, $R_{xx}$ is less sensitive
to the choice of the phases.  If the phases are zero, the main problem
is the upturn of $R_{xx}$ when temperature approaches to the SDW
transition temperature (see Fig.\ 4 of Ref.\ \cite{R_xx}).  The upturn
can be suppressed by selecting a small but finite value for the phases
(see the other curves in Fig.\ 4 of Ref.\ \cite{R_xx}) or by selecting
a greater value for $t_b'$ (see Fig.\ 6 of Ref.\ \cite{R_xx}).  The
second problem is the difference in the temperature scales $\alpha_T$
utilized to obtain the fits in Figs.\ \ref{fig:R_H} and
\ref{fig:R_xx}.  While the scale 1.3 for $R_{xx}$ is reasonable, the
value 2.7 for $R_H$ is too big.  However, the scale of $R_H$ is
controlled primarily by $t_b'$ (because $R_H^{(1)}=0$ if $t_b'=0$
\cite{R_H}), whereas the scale of $R_{xx}$ is controlled by $t_b$.  So
the agreement could be achieved by increasing $t_b'$ without
increasing $t_b$.  The increase of $t_b'$ could suppress the SDW
transition, which is present at the ambient pressure, but that could
be compensated by an increase in the interaction strength.

The space of model parameters is big, and there are opportunities to
achieve better agreement with the experiment by optimizing the choice
of parameters.  Our first try presented here appears to be promising.

\vspace{-\baselineskip}

\end{document}